\documentclass{llncs}

\usepackage[utf8]{inputenc}
\usepackage[T1]{fontenc}
\usepackage[final]{graphicx}
\usepackage[labelsep=period]{caption}
\usepackage[hyphens]{url}
\usepackage[english]{babel}
\usepackage{amsmath}
\usepackage{algorithm}
\usepackage{algorithmic}

\def\thebibliography#1{\section{References}\list
{[\arabic{enumi}]}{\settowidth\labelwidth{[#1]}\leftmargin\labelwidth
\advance\leftmargin\labelsep
\usecounter{enumi}}
\def\newblock{\hskip .11em plus .33em minus .07em}
\sloppy\clubpenalty4000\widowpenalty4000
\sfcode`\.=1000\relax}

\begin{document}

\title{Visual analytics in FCA-based clustering}

\author{
Yury Kashnitsky
}

\institute{
National Research University Higher School of Economics, Moscow, Russia
\email{ykashnitsky@hse.ru}
}

\maketitle

\begin{abstract}
Visual analytics is a subdomain of data analysis which combines both human and machine analytical abilities and is applied mostly in decision-making and data mining tasks. Triclustering, based on Formal Concept Analysis (FCA), was developed to detect groups of objects with similar properties under similar conditions. It is used in Social Network Analysis (SNA) and is a basis for certain types of recommender systems. The problem of triclustering algorithms is that they do not always produce meaningful clusters. This article describes a specific triclustering algorithm and a prototype of a visual analytics platform for working with obtained clusters. This tool is designed as a testing frameworkis and is intended to help an analyst to grasp the results of triclustering and recommender algorithms, and to make decisions on meaningfulness of certain triclusters and recommendations.

\vspace{1em}

\end{abstract}

\section{Introduction}
\label{intro}
\hspace{0.35cm} Classical Formal Concept Analysis (FCA) deals with data which describe a relationship between a set of objects and a set of attributes and provides methods to derive a concept hierarchy or formal ontology in them \cite{FCA}. FCA is a powerful tool for revealing dependencies in data and is commonly applied to data mining (in particular, text mining), machine learning, knowledge management, semantic webs, software development, and biology.

	As a natural extension of FCA, Triadic Concept Analysis (TCA) manages triadic data in a form of objects, their attributes, and conditions under which these objects have certain attributes \cite{TCA}.
A common example is a social network analysis with a context including users (objects), events they take part in (attributes) and interests (which might be regarded as conditions under which a user participates in a certain event).

	As the task of finding all concepts or triconcepts is computationally challenging, certain relaxations of these terms have been introduced: biclusters \cite{Bicluster} and triclusters \cite{Tricluster}. Here we address triclusters, i.e. combinations of sets of objects, their attributes, and conditions where not every object must have each attribute. Triclustering provides an output in the form of object clusters with
similar attributes under similar conditions. Therefore, it is applied to mining users with common interests, applicants with similar competences or books labelled by close tags \cite{vk_analysis}, \cite{adver_bicl}. Triclustering is also a basis for a certain type of recommender systems \cite{perfume_rec}, \cite{recbi}.

	Visual analytics is an increasingly popular branch
of Computer Science which combines both human and
computer qualities to solve a range of problems that might
lay beyond the power of man or machine separately.
Actually, it is a subdomain of data analysis focusing on
decision-making through data preprocessing, data mining
and interactive user interfaces. For instance, Siemens
PLM software allows developers to collect, process, visualize
report data in the 3D engineering environment, and
make real-time decisions in the process of developing new
vehicles. The same method is used in situational and
decision-making centres, in nuclear power energetics, and
in crime investigations.

	In this paper, we explore these topics and describe a framework which uses visual analytics to solve some problems in FCA.


\section{Visual analytics}
\subsection{Definition and specificity}
\hspace{0.35cm} Generalizing and selecting crucial aspects of various definitions of visual analytics \cite{VisAnKeim}, \cite{VisAnCook}, here we propose the following one:

\emph{Visual analytics} is a subdomain of data analysis focusing on analytical reasoning on the basis of interactive user interfaces 
in process of data mining, data preprocessing, knowledge representation, discovering dependencies, and decision-making.

Let us further consider core peculiarities of visual analytics and the tasks it is designed to solve: \cite{Kielman}

\begin{enumerate}
\item Visual analytics usually deals with complicated problems with big amounts of data
requiring both human and machine resources. 
\item The final goal of visual analysis is to enable users to obtain deep insight in problems to be solved which might include processing of large amounts of data from various sources. For this purpose visual analytics combines both human and technological resources. On one hand, data mining and statistics are the driving force of any automatic data analysis. On the other hand, human brain's aptitude for information perception and discovering dependencies in data complies to machine techniques and thus turns visual analytics into a promising sphere for further development.
\item In its development, visual analytics fosters in its turn the development of data mining, data representation and visualization, and analytical reporting.
\item Visual analytics also deals with human cognition, information perception, Computer Science, interactive and graphical design.
\item Visual analytics combines methods of information visualization and graphical data representation where visualization fosters human perception by the following means:

\begin{enumerate}
\item Enlarging data resources capacities makes user memorize less
\item Reducing search, such as by representing a large amount of data in small space
\item Enhancing recognition of patterns, such as when information is organized in space by its time relationships
\item Supporting easy relationship inference 
\item Monitoring large amounts of potential events
\item Providing techniques for dynamic data monitoring
\end{enumerate}
\end{enumerate}

\subsection{Siemems}
\hspace{0.35cm} Siemens uses visual analytics techniques in its product lifecycle management (PLM) software enabling developers to collect, process, visualize report data in the 3D engineering environment, and make real-time decisions in the process of developing new vehicles. \footnote{http://www.plm.automation.siemens.com}

\begin{figure}[!h]
\centering{\includegraphics[width=7.0 cm]{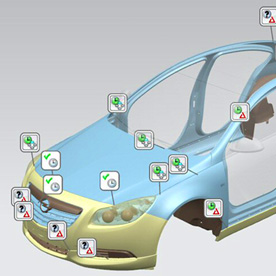}}
\caption{One of development stages with Siemens PLM Software}
\label{periodic}
\end{figure}

The crucial point is that this system allows real-time visual interaction. This speeds up the processes of testing production for meeting given criteria, and eliminating product quality problems. 

\subsection{Supernova modelling}
\hspace{0.35cm} A highly powerful implementation of visual analytics paradigm was fulfilled by astrophysicists in Terascale Supernova Initiative (TSI) project. \footnote{science.energy.gov/$\sim$/media/ascr/ascac/pdf/meetings/mar03/Mezzacappa.pdf} The goal of the project is to give scientists from various fields access to powerful computation resources in order to produce knowledge in the sphere of fundamental science. In particular, the question of supernova birth was studied which encompassed 3D turbulence, gravitation and magnetic field modelling. The scale of the investigation was impressive - the modelling resulted in terabytes of data. The analysis of such amount of data lays beyond human power but combining human and machine capabilities allowed to make some inferences from all the bulk of data.


\section{Formal Concept Analysis and OA-biclustering}
\label{section_FCA}

\subsection{Main definitions}

\hspace{0.35cm} A \textit{formal context} in FCA is a triple $K = (G, M, I)$ where $G$ is a set of objects, $M$ is a set of attributes, and the binary relation $I \subseteq G \times M$ shows which object possesses which attribute. $gIm$ denotes that object $g$ has attribute $m$. For subsets of objects and attributes $A \subseteq G$ and $B \subseteq M$ \textit{Galois operators} are defined as follows:
\begin{equation*} 
\begin{split}
A' = \{m \in M\ |\ gIm\ \forall g \in A \}, \\
B' = \{g \in G\ |\ gIm\ \forall m \in B\}. 
\end{split}
\end{equation*}

A pair $(A, B)$ such that $A \subset G, B \subset M , A' = B$ and $B' = A$, is called a \textit{formal concept} of a context $K$. The sets $A$ and $B$ are closed and called the \textit{extent} and the \textit{intent} of a formal concept $(A, B)$ respectively. For the set of objects $A$ the set of their common attributes $A'$ describes the similarity of objects of the set $A$ and the closed set $A''$ is a cluster of similar objects (with the set of common attributes $A’$).
 
The number of formal concepts of a context $K = (G, M, I)$ can be quite large ($2^{min\{|G|, |M|\}}$ in the worst case), and the problem of computing this number is \#P-complete \cite{hard}. There exist some ways to reduce the number of formal concepts, for instance, choosing concepts by stability, index or extent size \cite{concept_reduce}.
 
An alternative way is to make a relaxation of the definition of a formal concept. One of them is an OA-bicluster \cite{Bicluster}. \\ If $(g, m) \in I$ , then $(m' , g')$ is called an \textit{object-attribute bicluster} with the \textit{density} 
$$\rho(m', g') = \frac{|I \ \bigcap \ (m' \times g')|}{(|m'||g'|)}.$$
Bicluster density represents the percent of object-attribute pairs from the initial context in a certain bicluster.  

Here are the main properties of OA-biclusters:
\begin{enumerate}
\item For any bicluster $(A,B) \subseteq 2^G \times 2^M$ it is true that $0 \leq \rho(A, B) \leq 1$,
\item An OA-bicluster $(m',g')$ is a formal concept if $\rho = 1$,
\item If $(m',g')$ is a bicluster, then $(g'', g') \leq (m', m'')$.
\end{enumerate}

A bicluster $(A, B)$ is called \textit{dense} if its density is greater than a predefined minimum threshold, i.e. $\rho((A,B)) \geq \rho_{min}$. The above mentioned properties show that OA-biclusters differ from formal concepts since unit density is not required. Below follows an illustrative example for triconcepts and triclusters.


\section{Triadic FCA and OAC-triclustering}
\hspace{0.35cm} As a solution for three-way data in FCA, Triadic Concept Analysis (TCA) was introduced \cite{TCA}.

A triadic context $K = (G, M, B, I)$ consists of sets $G$ (objects), $M$ (attributes), $B$ (conditions), and ternary relation $I \subseteq G \times M \times B$. An incidence $(g, m, b) \in I$ shows that the object $g$ has the attribute $m$ under condition $b$.

We denote a triadic context by $(X_1, X_2, X_3, I)$. A triadic context $K = (X_1, X_2, X_3, I)$ gives rise to the following dyadic contexts: 
\begin{equation*} 
\begin{split}
K^{(1)} = (X_1 , X_2 \times X_3 , I^{(1)}), \\
K^{(2)} = (X_2 , X_3 \times X_1 , I^{(2)}), \\ 
K^{(3)} = (X_3 , X_1 \times X_2 , I^{(3)}),
\end{split}
\end{equation*}
where $gI^{(1)}(m, b) \Leftrightarrow mI^{(1)}(g, b) \Leftrightarrow bI^{(1)}(g, m) \Leftrightarrow (g, m, b) \in I$. 

The derivation operators (or prime operators) induced by $K^{(i)}$ are denoted by $(.)^{(i)}$. For each induced dyadic context we have two kinds of derivation operators. 
That is, for $\{i, j, k\} = \{1, 2, 3\}$ with $j < k$ and for $Z \subseteq X_i$ and
$W \subseteq X_j \times X_k$ , the (i)-derivation operators are defined by:

$Z \rightarrow Z^{(i)} = \{(x_j , x_k ) \in  X_j \times X_k  \ | \ x_i , x_j , x_k$ are related by $I$ for all $x_i \in Z\}$, 

$W \rightarrow W^{(i)} = \{x_i \in X_i \ | \ x_i , x_j , x_k $ are related by $I$ for all $(x_j , x_k ) \in W \}$

A \textit{triadic concept} of a triadic context $K = (G, M, B, I)$ is a triple $(A_1, A_2, A_3)$ of $A_1 \subseteq X_1$, $A_2 \subseteq X_2$, $A_3 \subseteq X_3$ such that for every $\{i, j, k\} = \{1,2,3\}$ with $j < k$ we have $A_i^{(i)} = (A_j \times A_k)$. \\
$A_1, A_2$ and $A_3$ are called the \textit{extent}, the \textit{intent} and the \textit{modus} of $(A_1, A_2, A_3)$.

A set $T = ((m,b)',(g,b)', (g,m)')$ for a triple $(g,m,b) \in I$ is called an \textit{OAC-tricluster} (or object-attribute-condition tricluster or just tricluster) based on prime operators.
Here 
\begin{equation*} 
\begin{split}
(g,m)' = \{ b \ | \ (g,m,b) \in I\},\\
(g,b)' = \{ m \ | \ (g,m,b) \in I\},\\
(m,b)' = \{ g \ | \ (g,m,b) \in I\}.
\end{split}
\end{equation*}

The \textit{density}  of a tricluster $(A, B, C)$ of a triadic context $K = (G, M, B, I)$ is
given by the fraction of all triples of $I$ in the tricluster, that is \\
$\rho(A,B,C) =
\frac{|I \bigcap A \times B \times C|}{|A||B||C|}$.

The tricluster $T = (A, B, C)$ is called \textit{dense} if its density is greater than a predefined minimum threshold, i.e. $\rho(T ) \geq \rho_{min}$.
Just similarly to biclusters, triclusters have the following properties: \\

\begin{enumerate}
\item For every triconcept $(A, B, C)$ of a triadic context $K = (G, M, B, I )$ with nonempty sets $A, B$ and $C$ we have $\rho(A, B, C) = 1$,
\item For every tricluster $(A, B, C)$ of a triadic context $K = (G, M, B, I)$ with nonempty sets $A, B$ and $C$ we have $0 \leq \rho(A, B, C) \leq 1$.
\end{enumerate}

\subsection{Example}
\hspace{0.35cm} Let us consider a sample context $K = (U, I, S, Y)$, where $U = \{$Ed, Leo, Max$\}$ is a set of users, $I = \{$soccer, hockey$\}$ --- their interests, $S = \{$soccer.com, nhl.com, fifa.com, hockeycanada.ca$\}$ --- sites they have added to bookmarks, $Y\subseteq U \times I \times S$ is a ternary relation between $U,I,S$ which can be expressed by Table \ref{example_context}: 

\begin{center}
	\begin{tabular}{p{2 cm} p{3 cm} p{2 cm}}
		\begin{tabular}{|l|l|l|}
		\hline
		& $i_1$ & $i_2$ \\\hline
		$u_1$ & x & x \\\hline
		$u_2$ & x & x\\\hline
		$u_3$ & x & x \\\hline
		\end{tabular}
		&
		\begin{tabular}{|l|l|l|l|l|}
		\hline
		& $s_1$ & $s_2$ & $s_3$ & $s_4$\\\hline
		$u_1$ & x & x & x & x \\\hline
		$u_2$ & x & x & x &\\\hline
		$u_3$ & x & x & x & x\\\hline
		\end{tabular}
		&
		\begin{tabular}{|l|l|l|l|l|}
		\hline
		& $s_1$ & $s_2$ & $s_3$ & $s_4$\\\hline
		$i_1$ & x &  & x &  \\\hline
		$i_2$ &  & x &  & x\\\hline
		\end{tabular}
	\end{tabular}
		\begin{table}[!h]
		\caption{Sample context. Designations: $u_1$ - Ed, $u_2$ - Leo, 				$u_3$ - Max, $i_1$ - soccer, $i_2$ - hockey, $s_1$ - 				soccer.com, $s_2$ - nhl.com, $s_3$ - fifa.com, $s_4$ - 				hockeycanada.ca.}
		\label{example_context}
		\end{table}
\end{center}

Here, generally, we have $|U||I||S| = 24$ triples to analyze. But actually, this number is reduced to $11$, as there are lots of void triples present. 
 
Actually, users Ed, Leo and Max share the same interests and almost the same sites (all the difference is that Leo has not bookmarked hockeycanada.ca). The idea of clustering here is presented by a tricluster
 $T = (\{u_1,u_2, u_3\},\{i_1,i_2\},$\\$\{s_1,s_2, s_3, s_4\})$ with density $\rho = 11 / 24 \cong 0.46$. \\
It is just one pattern to analyze instead of 11 in case of triples.


\section{Implemented algorithms}
\hspace{0.35cm} The algorithms, described below, were implemented in Python 2.7.3 on a 2-processor machine (Core i3-370M, 2.4 HGz) with 3.87 GB RAM. One can find a description of testing procedure for these algorithms in \cite{tricl_exp_comparison} and \cite{vis_anal_tricl}.

\subsection{OAC-prime triclustering algorithm}
\hspace{0.35cm} The hard core of the algorithm is quite simple: for all incidences $(g, m,b) \in I $ for a triadic context $K = (G,M,B,I)$ we build a tricluster $T = ((m,b)',(g,b)',(g,m)')$. If a tricluster is unique and its density exceeds a predefined minimum threshold then it is added to an array of triclusters. A pseudocode of algorithm for OAC-triclustering based on prime operators is presented below.  

\begin{algorithm}
\caption{OAC-triclustering based on prime operators}
\label{alg:tricluster}
\textbf{Input:} $K = (G,M,B)$ - tricontext, \\
$\rho_{min}$ - density threshold \\
\textbf{Output:} $Tdic = \{X_1,X_2,X_3\}$ --- a tricluster dictionary. $X_1 \subseteq G, X_2 \subseteq M, X_3 \subseteq B$
\begin{algorithmic}
\FOR{$(g,m,b) \in I$}
\STATE $T = ((m,b)',(g,b)',(g,m)')$ \\
\STATE $HashKey = hash(T)$
\IF{$HashKey \notin Tdic.keys()$ and $\rho(T) \geq \rho_{min}$}
\STATE $Tdic[hashKey] = T$
\ENDIF
\ENDFOR
\end{algorithmic}
\end{algorithm}

\subsection{Recommender algorithm based on triclustering}
\label{section_rec}

\begin{algorithm}
	\caption{Recommender algorithm}
	\label{alg:recommendation}
	\textbf{Input:} $K = (U, T, R, I)$ - tricontext, $Tr$ - a set of 			triclusters \\
	\textbf{Output:} $Tag_{rec}, Res_{rec}$ - sets of recommended tags and 			resources
	\begin{algorithmic}
	\FOR{$u \in U$}
	\FOR{i = 1,...,|Tr|}
	\STATE $sim_{u}({Tr}_i) = \frac{1}{2}(\frac{|R_u \cap R_{{Tr}_i}|}{|R_u \cup R_{{Tr}_i}|} + \frac{|T_u \cap T_{{Tr}_i}|}{|T_u \cup T_{{Tr}_i}|})$ 
	\STATE ${Tr}_{best}  = argmax(sim_{u}({Tr}_i))$
	\STATE ${Tag}_{rec}[i] = T_{{Tr}_{best}} \setminus T_u$ 
	\STATE ${Res}_{rec}[i] = R_{{Tr}_{best}} \setminus R_u$ 
	\ENDFOR
	\ENDFOR
	\end{algorithmic}
\end{algorithm}

The recommender algorithm applied to sets of a tricontext is analogous to the one described in \cite{perfume_rec}. It takes as an input a context of three sets (objects, attributes, conditions), and the set of triclusters obtained as a result of the OAC-prime triclustering algorithm. For each user among all triclusters the one most similar to triples with this user is selected. The similarity of triclusters and triples is defined by function $sim_u({Tr}_i)$. The algorithm returns sets $Tag_{rec}, Res_{rec}$ - tag and resource recommendations for all users. 


\section{The challenge and visual tricluster analysis framework}
\hspace{0.35cm} The challenge of the problem of triclustering (as of clustering on the whole) is to output meaningful, well-interpreted clusters. Actually, the term "meaningful" is not formally defined and is used by people to express their own subjective opinion on how well the task of clustering is solved, i.e. how similar the objects in same clusters are, how distant - in different ones, how it corresponds to real world problems etc. Therefore, here an expert opinion might be useful, and a prototype of a visual analytics framework, described below, provides visual feedback for expert, and gives him ability to explore clusters in details. 

\begin{figure}[!h]
\centering{\includegraphics[width=7.5 cm]{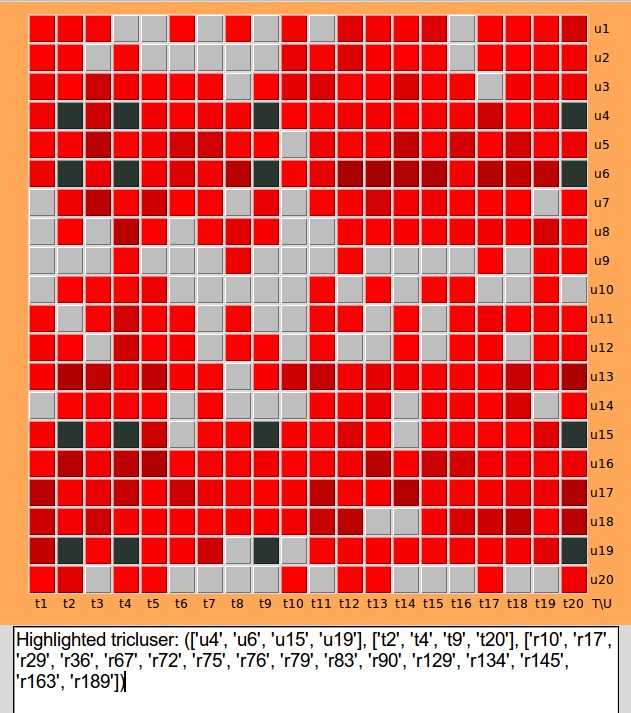}}
\caption{Highlighting a largest tricluster for a user-tag pair  $(u6, t4)$}
\label{fig:framework_largest_tricl}
\end{figure}

In figure \ref{fig:framework_largest_tricl}, we can see a map of triclusters produced by algorithm \ref{alg:tricluster} for a context of 20 users, 20 tags, and 200 resources. The map is projected on the User-Tag plane. The more a certain user-tag pair is presented in triclusters the darker the corresponding square. A user-tag pair $(u6, t12)$, for instance, is included in 73 triclusters (a dark red square) while $(u5, t9)$ - just in 1 (a red square), and no triclusters have a pair $(u9, t10)$ (a grey one). 

\begin{figure}[!h]
\centering{\includegraphics[width=7.5 cm]{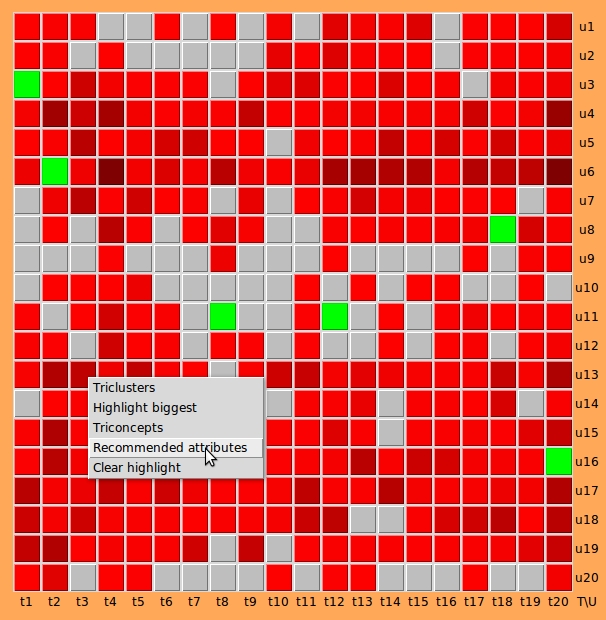}}
\caption{Recommended tags for several users}
\label{fig:framework}
\end{figure}

All triclusters including a certain user-tag pair can be listed by clicking on the "Triclusters" menu label. Similarly, triconcepts can be listed. One can also highlight the biggest tricluster with a certain user-tag pair or output all triclusters of the initial context ordered by density.  Moreover, through the "Recommend attributes" context menu option an analyst can depict the results of recommender algorithm for a certain user (in this case, to show recommended tags).

The tool is intended to help an analyst to grasp the results of triclustering and recommender algorithms, and to make decisions on meaningfulness of certain triclusters and recommendations. The map helps the expert to quickly detect the concentrated regions (dark squares) and visualize dense triclusters including the corresponding triples. Further, it helps to make the decision whether the selected dense tricluster is meaningful or not, i.e. if it really combines similar users, tags, and resources. 

\section{Further work}
There are several important issues to be regarded:

\begin{enumerate}
\item Limited human contribution: human contribution to triclustering in this visual analytics approach is limited and might only reach some hundreds of decisions on certain triclusters (less plausible, a thousand). Therefore, machine learning approach might help to learn the algorithm to classify meaningful clusters. The distance metric on triclusters should be carefully chosen.
\item Scalability: the issue of scalability is quite challenging in the described technique, and is to be solved. In current state, the application can support only contexts with one long dimension, for instance, a context of 20 users, 20 tags, and 400000 resources which can be projected onto a user-tag plane. One possible way to address the scalability issue is to perform preliminary clustering of objects, attributes, and conditions separately, and then choose representatives from each cluster. 
\item Extending the idea of a human-machine approach to other problems in FCA or data mining, such as exploring implications and association rules in order to find meaningful ones.
\end{enumerate}

\section{Conclusion}

\hspace{0.35cm} Visual analytics, as one of the flourishing domains of data analysis, can be useful in mining objects with similar attributes under similar conditions in a context of social network data. A special algorithm was developed for uniting such objects, attributes, and conditions in triclusters. The program framework under development is intended to graphically display the results of this algorithm and to empower an analyst to decide on the meaningfulness of clusters and tags or resources recommendations for objects.

\subsubsection*{Acknowledgements}
The author would like to thank his colleagues from Higher School of Economics Sergei Kuznetsov and Dmitry Ignatov for their well-timed advice and support during this work. He also expresses gratitude to Stanislav Klimenko from Moscow Institute of Physics and Technology for consulting in visual analytics.  

\bibliographystyle{splncs}

\end{document}